\documentclass[submission,copyright,creativecommons]{eptcs}
\providecommand{\event}{FMAS2023} 
\usepackage[utf8]{inputenc}
\usepackage{csquotes}
\usepackage{booktabs}
\usepackage{xcolor}
\usepackage{xspace}
\usepackage[inline]{enumitem}
\usepackage{sty/commands}
\usepackage{hyperref}
\usepackage{iftex}
\usepackage{lineno}
\usepackage{tcolorbox}
\usepackage{floatrow}
\usepackage[normalem]{ulem}
\ifpdf
  \usepackage{underscore}         
  \usepackage[T1]{fontenc}        
\else
  \usepackage{breakurl}           
\fi
\title{What to tell when? -- Information Provision as a Game}
\author{Astrid Rakow 
\quad\quad\quad\quad\quad Mehrnoush Hajnorouzi
\institute{Inst.\ of Systems Engineering for Future Mobility,\\
German Aerospace Center (DLR) e.V.}
\email{astrid.rakow@dlr.de\quad\quad\quad\quad mehrnoush.hajnorouzi@dlr.de}
\and
Akhila Bairy \qquad\qquad
\institute{Dept.\ of Computing Science,\\
Carl von Ossietzky University}
\email{\quad akhila.bairy@uni-oldenburg.de}
}
\newcommand{\titlerunning}{Information Provision as a Game}
\newcommand{\authorrunning}{A. Rakow, A. Bairy, M. Hajnorouzi}
\hypersetup{
  bookmarksnumbered,
  pdftitle    = {\titlerunning},
  pdfauthor   = {\authorrunning},
  pdfsubject  = {EPTCS},               
}
\begin{document}
\maketitle
\begin{abstract}
\emph{Constantly informing systems} (\CIS), which provide us with information over a long period of time, face a particular challenge in providing useful information:
Not only does a person's information base change but also their mood or abilities may change. 
An information provision strategy should hence take these aspects into account. 
In this paper, we describe our vision for supporting the design of \CIS. 
We envisage using psychological models of the human mind and emotions in a game of information provision.
The analysis of these games will give comparative insights into design variants.  
\end{abstract}
\section{Introduction}
Complex technical systems are deeply integrated into our daily lives.
They provide us with various services ranging from entertainment to safety-critical services. 
We use many of these systems over long periods of time and they even adapt to our needs. 
Such systems and their users exchange information:
On one hand, the system acquires user information to adapt. 
On the other hand, users often need information from the system \eg some explanation of the system's behaviour. 
When such an interaction between the system and its user extends over a longer period of time, an important question therefore arises: what content should be communicated to the user and when?
A few brief examples will illustrate the challenges of this question: 
\begin{enumerate*}[label={(\arabic*)}] 
\item An autonomous vehicle \AV initiates emergency braking due to a deer suddenly crossing the road.
\item The \AV slows down on the way to work due to a dumpy road.
\item The \AV slows down due to a carnival procession, the user is accompanied by his excited kids.
\end{enumerate*}
In these scenarios, the \AV faces the challenge of determining what content helps the user in the current context. 
It should factor in that the time span is very short in (1), that the driver has prior knowledge in (2), and the right level of detail is especially important, if the driver is occupied otherwise as in (3).

\begin{figure}[h]
\floatbox[{\capbeside\thisfloatsetup{capbesideposition={right,top},capbesidewidth=9cm}}]{figure}[\FBwidth]
{\caption{\CIS observes its environment and provides information to a human model (\HM). \HM perceives this information and observes its surrounding Env to choose its actions. The goal of \CIS is to enable \HM achieve its goals.}\label{fig:game}}
{\includegraphics[width=6.7cm]{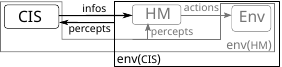}}
\end{figure}

In this paper, we argue that the design of a \emph{constantly informing system (\CIS)}%
\footnote{%
Self-explaining systems are also \CIS. A \CIS is not necessarily  self-explaining.}%
 can profit from a holistic game theoretic analysis that places a strong emphasis on the human aspect.
We propose to analyze the challenge of information provision \textit{(What content should be communicated to the user when?)} by methods of economics to determine which game is to be played (Sect.~\ref{sec:rightgame}). In the resulting game, we consider the human as a system of limited cognitive resources (Sect.~\ref{sec:analysis}). 
In order to do so, we propose to use models of the human mind and emotions from psychology, which we briefly survey in Sect.~\ref{sec:models}.
We outline how we envision integrating these models into games between a \CIS and its user (cf. Fig.~\ref{fig:game}).
Their analysis results will allow designers comparing variants of a \CIS under development.

\section{The Challenge of Information Provision Seen as a Game}\label{sec:rightgame}

Game theory, developed by von Neumann and Morgenstern \cite{vonNeumannMorgenstern1947GameTheory} in the mid-20th century, studies interactive decision-making where the outcome for each player depends on the actions of the other players \cite{Myerson91}.
It has been used in various domains such as economics, political science or
computer science. 
In the following, we discuss how the challenge of information provision can be considered a game.
By reinterpreting the economic analysis method PARTS~\cite{PARTS} for information provision, we outline the broader scale of this challenge and then narrow the focus down to the next steps of our research.
PARTS defines a list of guiding questions to analyze a situation as a game: 
the \textbf{p}layers \expl{(who is active? who has goals?)}, 
\textbf{a}dded value \expl{(what does player X gain by Y's participation?)}, 
\textbf{r}ules \expl{(what rules do constrain the players? who can change the rules?)}, 
\textbf{t}actics \expl{(how can the players' perception of the game be changed?)} and 
\textbf{s}cope \expl{(what are the boundaries of the game?)}.
PARTS is only a first step towards the definition of a game.
The concrete game depends on the characteristics of the application domain, the information available to build a model of the application domain, and the goal of the analysis. 

We want to analyze how the \emph{\enquote{usefulness of a \CIS}} can be optimized. 
\emph{Useful} information is relevant information\footnote{
Here, relevant information for the \CIS's customer is the minimal information that they need to accomplish their goal \cite{relevanceTR}.
} that the customer can \emph{process and use} to accomplish their goals, i.e.\ the right content at the right time.
It is part of our future work to concretize this notion. 
Table \ref{tab:roles} summarizes how we interpret the \textbf{p}layers' roles in an information provision setting. 
While the \textbf{a}dded values that the players will bring to the game vary depending on the system considered, we can identify two main types of \textbf{r}ules as shown in Table \ref{tab:rules}.
\textbf{T}actics refer to actions that a player can take to change how other players perceive the game, \ie
to change their belief in  who the players are, what their added values and what the rules are. 
The \textbf{s}cope of \emph{our} game is a \CIS that is a subsystem of an \AV and serves a single human (the driver) who is captured via a human model \HM. 
For a start, we use a game of this very limited scope  to investigate how to formally include human capabilities in the design of a \CIS.
\begin{table}[h]
\small
\begin{tabular}{l p{13cm}}
\toprule
Customers & all recipients of information (humans or technical systems)\\
Suppliers & suppliers of information (sensors and other \CIS)\\
        & suppliers of computing services (hardware resources)\\
Competitors & systems that render the \CIS's information redundant (other \CIS, human perception)\\
        & processes that use the same computing services or human resources \expl{(\eg music system)}\\
Complementors & other \CIS whose information enhances the \CIS' information \expl{(\eg traffic warnings plus detour)}\\
\bottomrule
\end{tabular}
\caption{Interpretation of player roles in the \enquote{information provision} game}\label{tab:roles}
\normalsize
\end{table}
\begin{table}[h]
\small
\begin{tabular}{lp{13cm}}
\toprule
 guiding rules & a player is obliged to follow these rules, but can break them at will (\eg ethical, legal rules)\\
 constraint rules & a player cannot break these rules; the rules can only be changed by a third party (resulting from the technical implementation)\\
\bottomrule
\end{tabular}
\caption{Types of rules in the \enquote{information provision} game}\label{tab:rules}
\normalsize
\end{table}%

As the above expos\'e shows, information provision is an interdisciplinary challenge where humans  and technical systems interact -- many  \CIS are embedded into a human cyber-physical system of systems.
Legal and ethical aspects, as well as human capabilities, influence what a \CIS should provide as useful information. 
Whether information is useful strongly depends on whether the information is relevant \emph{and} on whether humans can make use of it.%
%

%
While the above was concerned with analyzing the application domain, we will now discuss what game \game in the \enquote{catalog} of well-studied games and respective analysis methods is appropriate for our analysis goal. 
We will take a closer look at the characteristics of the application domain and the goal of the analysis -- keeping in mind  what aspects are most important and which aspects might be omitted.
In a nutshell, the goal of \game's analysis is to determine which content our \CIS should provide to a \HM in what context. 
This analysis is meant to be done at design time in order to derive requirements on the information provision and to compare different design alternatives.
Our \CIS controls what information content it provides when to \HM. 
We discuss what kinds of games are appropriate using the following guiding questions based on the taxonomy of games as developed in \cite{MouraHutchinson}: \emph{Is \CIS acting strategically? What is the relation between \HM and \CIS? Who knows what?} 

\mypara{Is \CIS acting strategically?} 
%
We will study a \CIS, that \emph{chooses rationally} the content that seems the most useful. We hence will develop a classical game rather than an evolutionary game.

\mypara{What is the relation between \HM and \CIS? Should we use cooperative or non-cooperative game theory?}
%
In cooperative game theory (\CGT), the players cooperate with each other and form coalitions to obtain maximum reward/payoff.

In non-cooperative game theory (\NCGT), the players make decisions independently and each player aims to maximise their own payoff.
We assume that \HM has a prioritized list of goals. 
\CIS has only the goal to be helpful within its constraints thus acting subordinately \footnote{It is out of this paper's scope to discuss the case when goals of \HM contradict the constraints of \CIS.}.

In such a subordinate-dominant relationship, both \CGT as well as \NCGT could be a fit.
In order to apply \NCGT we envision to model \HM as always greedily accepting all information that it can in its current \emph{mode}; \expl{\Eg if \HM is tired, less information per time is accepted.}  
We thereby encode a kind of cooperation into \HM's modes, 
so that it is not playing against \CIS by maliciously rejecting all offered information.
To derive these modes and the dynamics of mode changes we plan to use psychological models of the human mind and emotions. 
We will give a brief overview of these in Sect.~\ref{sec:models}.
In Sect.~\ref{sec:analysis} we will give technical details on how we envision using such psychological models to derive the \HM for our game.
%

\mypara{Who knows what?} 
Both \CIS and \HM receive information from the environment. 
In particular, \HM receives information from our \CIS within its car but also via its five senses that are competing with \CIS.
\CIS also does not exactly know how \HM reacts, as it has no precise means to derive the state and reactions of \HM.
From \CIS's point of view \HM's reactions expose some degree of randomness.
\HM is a system of bounded resources and will usually not (have the resources to) think much about the beliefs of any support systems in its car.
We hence will define an epistemic game, where \CIS has a belief on \HM's belief but \HM does not build beliefs on \CIS beliefs.

\CIS knows what the possible inputs of \HM are, but assumes that these channels are unreliable.
That is, we assume that \CIS can roughly predict what information \HM is able to receive via its senses.
\CIS moreover knows what actions \HM can choose (what a driver can do). 

To us, the central challenge in the information provision game is capturing the human factors.
We envision to derive instantiations of \HM from models of the human mind and emotions. 
These instantiations are then \enquote{plugged} into our game.
We then check whether our \CIS can inform \HM, \ie whether a strategy exists, such that \HM achieves its goals.

\section{Models of Human Mind} \label{sec:models}
For an automated system to provide useful information, it needs a human behaviour model to determine the impact of the content and the optimal time for providing the content.
Well-crafted explanations can reassure users and increase alertness~\cite{Koo.2016}. 
The impact of percepts and its timing --or more generally the dynamics of the human mind-- is captured by cognitive architectures. 
These are well-established frameworks developed in the field of cognitive psychology \cite{newell90,anderson98}.
They simulate how knowledge is processed, stored, and utilized in response to external stimuli, perception, and knowledge \cite{peebles10}.
There are many cognitive architectures with different foci, like ICON FLUX\cite{nivel2007ikon}, CASCaS \cite{Wortelen13}, PRIM\cite{PRIM}, but none of them captures the human holistically.
In addition to modelling core cognitive functions (such as perception, learning, memory, etc.), a few cognitive architectures can also capture the interplay of emotions and behaviour.
These architectures generally adopt two primary approaches to model emotion-induced behaviours: altering architecture structures (\eg modifying memory modules), or modifying the processing parameters (\eg adjusting retrieval delay).
Emotions act as conduits to human behaviour, significantly impacting decisions and influencing the daily choices of individuals \cite{BecharaDD.2000}.
A change in the person-environment relationship \cite{lazarus91} may trigger emotional shifts, which may change this relationship in future.

Over time, different theories of emotion have evolved to explain the functioning of human emotions \cite{Oatley87,lazarus82,izard92}. 
The cognitive appraisal theory is one such theory. 
According to this theory, emotions are influenced by individuals' cognitive evaluations and interpretations of situations rather than by the situations themselves. 
Many computational models are built on this theory \cite{MarsellaG-EMA.2009, Becker-AsanoW.2010, LinSBZ-EmoCog.2011}.
Implementation of the appraisal process within the context of cognitive architectures is an ongoing effort. 
MAMID~\cite{hudlicka02} is one example.
It processes incoming stimuli through various modules. 
The affect appraiser module derives emotional states.
The resulting affective state influences the cognition processes by adjusting the rules related to goal selection, action determination, module speed and capacity, and mental construct prioritization.

The appraisal process provides an assessment of the current situation, influenced by various parameters, such as individual history, personality, and current affective state. 
These parameters can be described by modelling the effects of transient states (emotions) and personality traits (characteristics). 

To depict how the current situation affects human behavior, \cite{zacharias96} incorporated a situation assessor module into their proposed SAMPLE\footnote{Situation Awareness Model for Pilot-in-the-Loop Evaluation} pilot model architecture. 
This module translates situation descriptions into types that trigger specific responses.
The module employs belief nets to hold required knowledge, with nodes denoting particular features and events, and links illustrating causal and correlational connections between them. 
Furthermore, weighted links denote transition probabilities.

Appraisal theory is also utilised in the field of affective computing to enhance systems' abilities to perceive and generate human emotions.
Affective computing seeks to make computers more human-like by enabling them to recognise, interpret, predict, and respond to human emotions.
While few models effectively integrate emotion generation and its impact on cognitive processes, EmoCog \cite{LinSBZ-EmoCog.2011} stands out. 
This work adeptly fuses diverse models of computational emotions and cognitive architectures into a unified model.
In cognitive architectures, emotions can be divided into two parts: emotion effects and emotion generation \cite{LinSBZ-EmoCog.2011}.
Emotion effects consider how emotions might influence cognitive processing, while emotion generation explores how cognitive processes contribute to the generation of emotion. 
EmoCog, developed by Lin et al., also has primary and secondary appraisals similar to WASABI, developed by Becker-Asano and Wachsmuth \cite{Becker-AsanoW.2010}. 
However, EmoCog's secondary appraisal is more versatile, explaining a broader range of emotions than WASABI. The model includes an appraisal module that adjusts arousal and valence values in short-term memory nodes.

A psychological model of a human can also be achieved by their attention assessment. 
Wickens et al. developed SEEV \footnote{Salience, Effort, Expectancy, and Value} model to quantify and predict a person's attention level across various areas of interest in a given situation \cite{Wickens.2001}.
Initially intended for pilots' attention prediction, \cite{Horrey.2006} and \cite{Wortelen:Phdthesis} later adapted SEEV to gauge driver attention in road traffic. 
Leveraging SEEV, Bairy and Fr{\"a}nzle crafted a model that predicts the optimal time to deliver explanations to drivers based on their attention level \cite{BairyF.2023}.

\section{Psychological Models in A Game}\label{sec:analysis}
In a nutshell, we want to define a light-weight game as sketched in Fig.~\ref{fig:game}. 
\CIS plays against its environment env\small(CIS)\normalsize, which contains the human captured via \HM.
As discussed in Sect.~\ref{sec:rightgame} we encode into \HM how a human reacts so that we do not need to consider the human as a separate player.
\CIS's actions are providing information to \HM. 
The \CIS model can reflect that information can only be provided if \CIS has the information in its information base and if it has the required computing resources and time to retrieve it.
Strategy synthesis will then tell a designer whether \CIS can provide information such that \HM achieves its goals
\footnote{This list of goals varies with the considered scenario but we assume that the goal of being safe has highest priority while the goal of feeling well has less priority.}.
A designer can plug in different variants of \CIS and compare how well they perform in terms of the goals that can be achieved.

\begin{figure}[h]
\floatbox[{\capbeside\thisfloatsetup{capbesideposition={right,top},capbesidewidth=8cm}}]{figure}[\FBwidth]
{\caption{\HM perceives its environment and chooses actions. During a scenario \HM changes its mode, reflecting a change in its cognitive resources, emotions, and information.}\label{fig:humanInLoop}}
{\includegraphics[width=7.2cm]{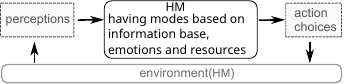}}
\end{figure}

We aim for a \HM model that encodes how a human reacts to the provided information or other environmental perceptions.
We assume that \HM reacts internally, \ie emotionally, by consuming cognitive resources or by changing its information base.
This internal change might cause a mode change of \HM. In the new mode \HM chooses a different strategy to reach its goals.
We thus think of \HM as sketched in Fig.~\ref{fig:humanInLoop}.
We want to apply strategy synthesis to examine the information provision game of \CIS (cf. ~Fig.~\ref{fig:game}) that uses such a \HM as part of its environment model. 
There, a dominant strategy would be the best strategy of \CIS to provide the \HM with the necessary information for \HM to achieves its goals.

A central idea of our approach is to consider \HM as a resource-bounded system that reacts mode-specifically to the provided information content.
How \HM reacts will be derived from psychological models that are instantiated for the considered scenario.
We consider the central challenges of making these models usable in our game to be:
\emph{attaining a formal representation of psychological models}, 
\emph{reducing their complexity}, 
\emph{combining different psychological theories}, and 
\emph{interpreting the results}.

\begin{figure}[htbp] 
\centering
\includegraphics[width=0.95\textwidth]{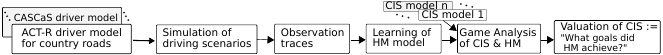}
\caption{From cognitive architectures to a comparison of \CIS variants.}\label{fig:games}
\end{figure}

As outlined in Sect.~\ref{sec:models}, there is a broad foundation of psychological theories that are shedding light on different aspects of how humans react to provided information.
The level of precision of these theories varies heavily as well as their level of formalization, ranging from informal over computational to formal. 
As Reisenzein puts it \enquote{Theoretical fragmentation and a comparative lack of
precision are [\ldots] characteristic of psychology in general}~\cite[p.248]{Reisenzein.2013}.
We focus on executable cognitive architectures as a starting point for our work, as they have an appropriate level of precision. 
But to use such architectures within a formal game, we need a formal representation of them.
The approach of Langenfeld {\etal}~\cite{Langenfeld19} for deriving timed automata for an ACT-R instance promises a detailed study of the human reactions as captured in ACT-R.
Since our precursor work \cite{BairyF.2023,Hajnorouzi23} showed that the resulting models easily get too complex for our needs, we want to derive simplified human models $\HM_i$ from cognitive architectures by automata learning techniques \cite{Aichernig2018}.
Automata learning techniques have been applied in many areas, such as speech recognition, software development, and computational biology~\cite{Vaandrager}.
In passive learning, the learning algorithm observes the inputs and the corresponding outputs of the system that is to be learned.
Active learning, on the other hand, adopts an interactive approach where the learner also actively asks strategic queries and observes the response. 
%
%
%
%
We want to use modes as a means for guiding how the human models get simplified.
Our idea is that a mode roughly specifies how \HM reacts, \ie what kind of strategy \HM will apply.
We think this might be done in a similar way to \cite{Ghosh2018}.
There Gosh and Verbrugge captured the kinds of reasoning strategies of player in a marble drop game as logical formulas, translated them into cognitive models and executed them in the PRIMs architecture \cite{PRIM}.
Instead of player types as in \cite{Ghosh2018} we think of player modes along the line \expl{\enquote{A tired driver reacts slowly. A driver gets tired when they are bored over a period of time.}} or \expl{\enquote{A driver that already has to keep seven facts in mind, will not easily remember five more facts.}\cite{miller.1956}}
We then want to apply automata learning to the cognitive architecture to derive a simplified model \HM.
There the choice of observable propositions will be an important influence on whether meaningful modes can be derived. 
Since the different architectures have different foci, we plan to use several architectures.
Given we would have $n$ appropriately light-weight models $\HM_i$ --derived from different cognitive architectures or manually designed--, we build $n$ games $\game_i$, $1\leq i\leq n$ (cf. Fig.~\ref{fig:games}).
The performance scores of the same \CIS $C_j$ in the different games $\game_i$ --measured e.g. in terms of \HM reaching its goals-- will lead to a $n$-tuple $t_j$ of scores for $C_j$.
The comparison of two \CIS variations, $C_j$ with performance $t_j$ and $C_k$ with performance $t_k$, will help the designer to get early on insight into how well the \CIS is tailored to human needs.
But the comparison of $C_j$ and $C_k$ has to be done with care,   
as the meaning of the $\HM_i$ models has to be taken into account.
This comparison gets especially difficult if the scores relate to interdependent aspects.
What if $\HM_h$ suffers from a high workload while $\HM_i$ is in a good mood?
Due to the coarse granularity of models that we want to consider, we expect that such cases will often occur, so that plausibility checks and if necessary refining analysis has to be done.
To this end, we think of spotlight approaches where the learner gets more fine-grained observations where necessary.
So, if this approach is in use, a collection of appropriate $\HM_i$ models could be evolved over time.
%

\section{Conclusion}
In this paper, we outlined our vision of how  game theory can be used to support the development of constantly informing systems (\CIS), taking into account the human factor. 
Our vision is to use models of the human mind and emotions from psychology to derive coarse formal human models, $\HM_i$ that can be used in a game between the \CIS and the human captured by $\HM_i$. 
The main challenges of our vision are attaining a formal representation of psychological models, reducing their complexity, combining different psychological theories, and interpreting the results.
Nevertheless, we hope to obtain at least comparative results  that can guide the early design of \CIS in tailoring to the human needs.
\appendix
\section{Acknowledgements}
This research has been supported by the German Research Council (DFG) in the PIRE Projects SD-SSCPS and ISCE-ACPS under grant no. DA 206/11-1, by RTG SEAS, and by the German Federal Ministry of Education and Research (BMBF) within the project "ASIMOV-D" under grant agreement No. 01IS21022G, based on a decision of the German Bundestag.

\section{Abbreviations}

\begin{tabular}{l l}
    \AV & Autonomous Vehicle \\
    \CIS & Constantly Informing System \\
    \CGT & Cooperative Game Theory\\
    \game & Game \\
    \HM & Human Model \\
    \NCGT & Non-Cooperative Game Theory\\
\end{tabular}

\bibliographystyle{eptcs}
\bibliography{reference}
\end{document}